# Characterization of High Temperature Calibration Bath through Stability and Uniformity Tests with Data Acquisition using Standard Platinum Resistance Thermometer and Precision Multimeter

**Arlene G. Estacio[1], Vilma C. Pagtalunan[2], Rio S. Pagtalunan[3], Ira C. Valenzuela[4], Lean Karlo S. Tolentino[5], and Jennifer C. Dela Cruz[6]**

[1,6]School of Electrical, Electronics, and Computer Engineering, Mapúa University, Philippines,
jennycdc69@gmail.com

[1,6]School of Graduate Studies, Mapúa University, Philippines

[1]Metals Industry Research and Development Center, Department of Science and Technology, Philippines,
agestacio@mirdc.dost.gov.ph

[2]Department of Electrical Engineering, Technological University of the Philippines, Philippines,
vcpagtalunan@yahoo.com

[3]Metals Industry Research and Development Center, Department of Science and Technology, Philippines,
rspagtalunan@mirdc.dost.gov.ph

[4,5]Department of Electronics Engineering, Technological University of the Philippines, Philippines

[4]University Research and Development Services Office, Technological University of the Philippines,
ira_valenzuela@tup.edu.ph

[5]University Extension Services Office, Technological University of the Philippines, Philippines,
leankarlo_tolentino@tup.edu.ph

## ABSTRACT

Calibration is one area in measurement where our country is still in the process of improvement. The absence of confirmed approaches and actions for identifying the features of calibration baths by numerous laboratories disallowed them to assess more the finest measurement ability and develop their accurateness. In recent advances in technology, there is now a way of providing the best measurement capability in temperature calibration by providing validated methods in characterizing calibration baths that contributes to the uncertainty of measurement of the calibrated thermometer and is therefore important to measure the extent of its contribution to the final uncertainty. This study adapted the process from obtainable and current methods. Platinum Resistance Thermometer (PRT-Isotech) and Standard Platinum Resistance Thermometer (SPRT-Hart) which are interfaced to Precision Digital Multimeters were used for measurements. Quite a few measurements were made at varying time durations implemented on several known positions in an orderly measurement form which enclosed the whole workspace. The uniformities and stabilities of the calibration bath were computed by getting the range of the Minimum Difference with the Maximum Difference taken at every established temperature point. Outcomes of measurements exhibited that the calibration bath stability at every temperature setting are steadier at lesser temperatures and have a tendency to rise as the temperature approaches 250ºC.

The liquid bath was further unvarying at lesser temperatures. The variances found were insignificantly distributed since the Standard Deviation ranged only from 0.005 to 0.013. It indicates that the local calibration laboratories can also perform the tests for stability and uniformity on a calibration bath at an accuracy level comparable to national laboratories since the measured values obtained were within the values obtained in similar tests conducted in a national laboratory abroad and have very small standard deviations.

**Key words :** characterization, liquid calibration bath, measurement, stability, uniformity

## 1. INTRODUCTION

Different setups were established for some calibration services outside the Philippines. These were not easily replicated by almost all calibration labs since the at one time comprehensive approaches and actions in facilitating the tests for calibration enclosures/bath in terms of their stability and uniformity were not provided. The absence of tested approaches and actions in identifying the faces of calibration baths by almost all home-grown calibration labs disallowed them to assess more their finest measurement ability and develop their accurateness [1-3].

This paper aims to identify the characteristics of this Liquid Calibration Bath at higher ambient temperatures by means of validated technique and process of the Metals Industry Research and Development Center (MIRDC) for measuring stability and uniformity. This objective was achieved, by (1) carrying out stability test by identifying the highest difference in temperature at a specified temperature set point and period, (2) handling uniformity test by identifying the highest variation in temperature among stated locations in the Liquid Calibration Bath, and (3) establishing a preliminary curve fitting model stating the performance of stability of Liquid Calibration Bath. The results of this study offer assistance to the home-grown facilities to accomplish at equivalence with labs outside the country offering comparable calibration services.  This provided more responsiveness to the students of the significance of calibration in this area of measurement, and finally, future endeavors in this field of metrology [1] can be explored by a university for a possible specialization in the undergraduate and graduate programs in the field of engineering and technology.

This study, though, is restricted on the usage of overflowing and stirred-liquid calibration bath. Different stirred-liquid baths' calibration chamber can put up several temperatures extending from -50°C to +600°C with regards to the kind of liquid utilized. Earlier test method initially was on low temperature range, yet, because of increasing temperature differences when one increased the temperature settings, it was noted that there were changes in the behavior of the liquid used such as bubbles created in the calibration bath, therefore, fixed points of test were done at its highest temperature of 250 ℃.

It limited the settings in temperature during test from 150°C to 250°C and applied one kind of liquid all over the measurement process and at minimum $23 \pm 3°C$ and $55 \pm 15\%RH$. The calibration bath utilized silicone oil. Measurements were made on several determined locations in an organized measurement form that consists of the whole workspace of the Liquid Calibration Bath [2].  The experimentation proper was performed in Instrumentation Laboratory of MIRDC located at Bicutan, Taguig City, Philippines.

The results of this paper align with the notion of traceability in measurement in which devices for measurement are adjusted by means of advanced accurateness standards which are noticeable to every country's National Laboratories [1], [4] as shown in Figure 1 below. Illustrating the behavior of calibration bath's uniformity and stability supports to intensify accurateness in calibration that improved uncertainty of measurement.

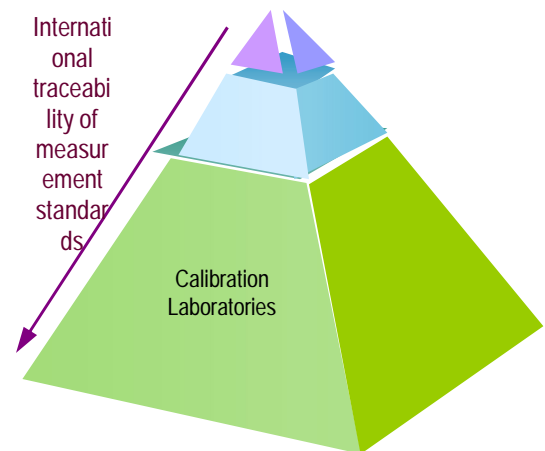

**Figure 1:** Traceability of measurements standards through laboratories [4]

Most of the calibration labs practice variable-temperature enclosure form since thermometers are adjusted by comparison scheme [4], as shown in Figure 2. For the said setup, thermometers are calibrated against with the other calibrated thermometer. The other calibrated thermometer was compared with one more adjusted thermometer of better accurateness in a calibrations chain that should eventually be traceable prior to a main standard thermometer [2], [5].

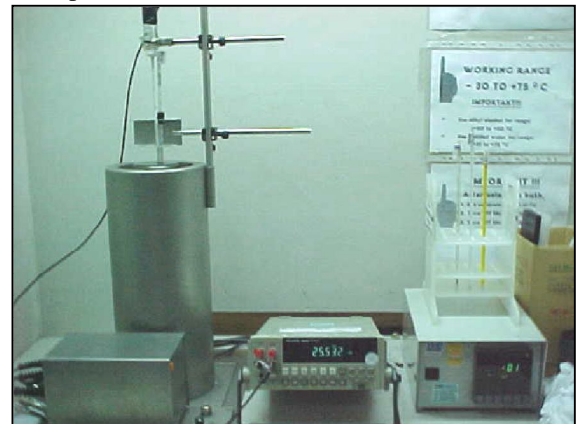

**Figure 2:** Comparison Method of Thermometers' Calibration Set-up

## 2. MATERIALS AND METHODS

The measurement methods took the procedure from accessible and current methods, e.g. MIRDC, [6-24] but generally, they were implemented by SPRT-Hart and PRT–Isotech thermometers which are interfaced to the precision Digital Multimeter. The said two thermometers were both submerged at the middle location for the testing of stability. Test of uniformity, instead, was executed at varying submersion lengths of 75mm, 150mm, 200mm, and 250mm while the temperature test points were secured at 150°C, 175°C, 200°C, 225°C, and 250°C. The SPRT was utilized to create the point of reference by submerging it at a depth of 200mm. It is made fixed at the calibration bath's halfway point.  The PRT was utilized for temperature measurement at varying test points. Careful attention during setup were done to guarantee that the liquid bath is occupied with silicone oil up to a level that will run-off from the calibration chamber's inner tubing [6]. The whole setup could warmup for 30 minutes before test point setting. After it was warmed and set up, the 1st test point, 150°C, was set on the calibration bath. Its stabilization period was observed. Several calibration bath models have dissimilar period of stabilization from ambient temperature to the "set point" temperature. In this paper, the used bath observed the requirement from the manual of the manufacturer. After the stabilization period, it was stabilized for 5 additional minutes before the 1st recorded digital multimeter reading. Measurements were made at varying periods.

The instruments as listed in Tables 1 and 2 below are the equipment utilized through the test.

**Table 1:** List of instrument used

| Description | Code No. | Brand/Model |
|---|---|---|
| Calibration Enclosure | T049 | ASL CB15-45e Calibration Liquid Bath |
| SPRT1 (Reference, Fixed) | T038 | Hart 5699 SPRT |
| SPRT2 (Test, Movable) | T033 | Isotech 670SH PRT |

**Table 2:** List of equipment used

| Indicator Used | Code No. | Setting |
|---|---|---|
| Agilent 34401A DMM | E050 | 4-wire, 100W |
| Fluke 8842A DMM | T038 | 4-wire, 100W |

A comprehensive procedural guide on the gathering of data was observed as specified in [4] (pp. 44-50).

## 3. RESULTS AND DISCUSSION

### 3.1 Result of Stability Test for 150°C to 250°C

Calibration bath stability at every temperature setting have a tendency to rise as the temperature approaches 250°C. As shown in Figures 3(a) to (e) and Figure 4, the stabilities of the calibration bath were computed by getting the Maximum Difference to Minimum Difference range taken at every temperature point, e.g. 150°C, 175°C. The variances found were insignificantly distributed since the Standard Deviation value is within 0.005 to 0.013. This specifies that the determined values are lower than the suitable limit of 0.01 to 0.05 which was performed in Australia's CSIRO where a similar study was performed [2].

Measured temperatures at lesser set point, e.g. -30°C, -20°C, 0 °C are more stable than with greater set points which similarly adapts to the measurement results done by [4] for setting which are less than 30°C.

The noted temperatures for analyzing stability were observed at a height of 200mm from the calibration bath surface. It is still uniform inside the uniformity test's results range for both horizontal and vertical analyses. Refer also to 3.2.

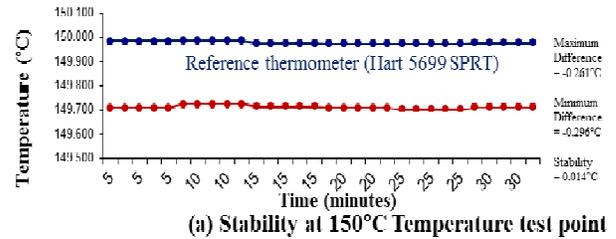

**(a) Stability at 150°C Temperature test point**

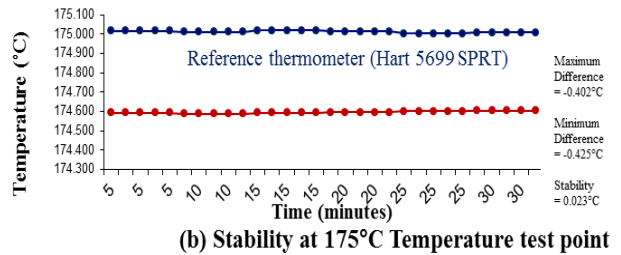

**(b) Stability at 175°C Temperature test point**

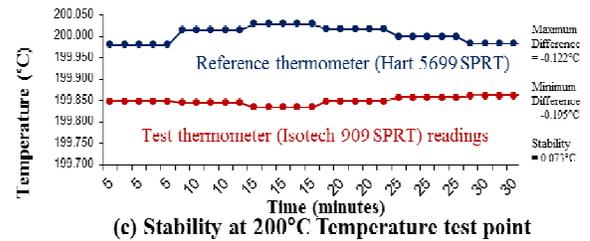

**(c) Stability at 200°C Temperature test point**

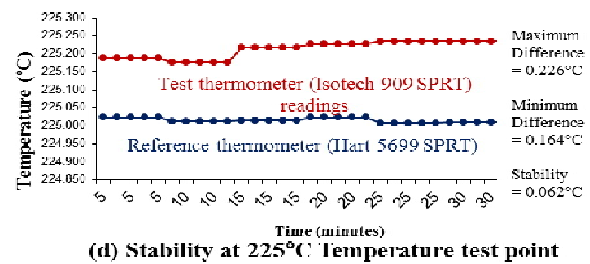

**(d) Stability at 225°C Temperature test point**

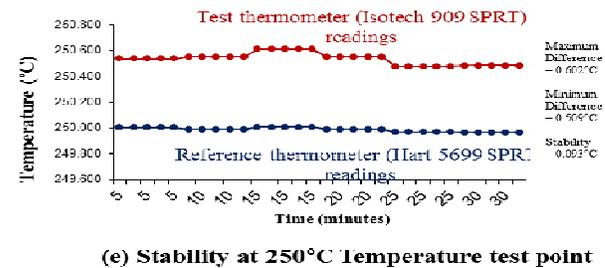

**(e) Stability at 250°C Temperature test point**

**Figure 3:** (a) to (e) Calibration bath's stability at various temperature settings

### 3.2 Result of Uniformity Test

Inside the calibration bath, he temperature's uniformity is largely caused by the liquid's motion. Irregular temperature may occur in the circulating system at several points [2].

When the liquid flow rate is improved, the bath's uniformity may be increased. Though, it is not recommended because amount of stirring must be limited to avoid overheating. Hence, the following part of this study is the assessment of the calibration bath's uniformity by means of horizontal and vertical positions.

### A. Uniformity test with regard to horizontal position

From the documented measurements gathered in the measurements results, the deviation of reading at every position, center, front, left, back, and right relative to the designated reference position was determined. In this part, the horizontal position illustrates the temperature uniformity behavior at different temperature set points, i.e., 150°C, 175°C, 200°C, 275°C, and 250°C. The differences of readings of the two SPRT's were determined. Taking into consideration that the Hart SPRT was fixed at the center position and Isotech PRT was placed at each designated positions. The differences were computed for each temperature setting. Using the values of the difference of readings with respect to horizontal position, i.e. center, front left, back and right positions in the calibration bath, a graph was generated to show the calibration bath's uniformity. Figure 4 shows the bath's uniformity with respect to horizontal position.

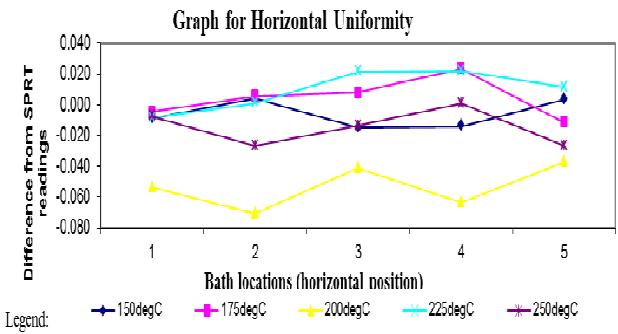

**Figure 4:** Result of Horizontal Uniformity

### B. Uniformity test with regard to vertical position

From the noted measurements gathered in the measurements results, the deviation of readings at each length of immersion relative to the designated reference length of immersion was determined. In this part, the vertical position illustrates the behavior at various immersion length, i.e. at 75, 150, 200, and 250mm in the bath for different locations, i.e. center, front, back, left, and right. This was computed for each temperature setting. The difference of readings of the two SPRT's was determined. Taking into consideration that the Hart SPRT was fixed at the center position and Isotech PRT was immersed at 75, 150, 200, and 250mm immersion lengths at each designated positions. This was computed for each temperature setting. Using the values of the difference of readings with respect to various immersion lengths or vertical position in the calibration bath, a graph was generated to show the calibration bath's uniformity. Figure 5 shows the uniformity of the bath with respect to vertical position.

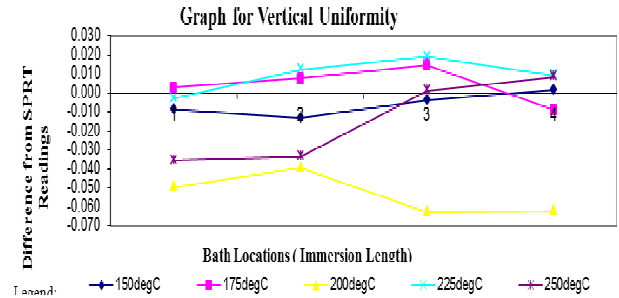

**Figure 5:** Result of Vertical Uniformity

## 4. CONCLUSION AND RECOMMENDATION

It is possible to say that the local calibration laboratories can also perform the stability and uniformity tests on a calibration bath at an accuracy level comparable to national laboratories since the measured values obtained based on the procedure outlined were within the values obtained in similar tests conducted in a national laboratory abroad and have very small standard deviations. After series of measurements done on the calibration bath's stability and uniformity and the analyses of gathered data, it is also concluded that the method and procedure can establish the characterization of a calibration bath controlled by temperature.

This further concludes that the method for testing implemented in this paper matches the industry's local requirement for accuracy. When the test temperature setting reaches 175°C, bubbles were observed to appear in the calibration bath. Presence of bubbles affect the calibration bath's uniformity and stability as this influence the homogeneity of the silicone oil.

Future studies include calibration baths which will be characterized using different liquids, e.g. ethyl alcohol and distilled water and development and implementation of software for high temperature calibration bath testing [25]. Likewise, further technique and process validation through assessment of measurements' uncertainty and En-value test can be done to broaden the applicability of the said characterization technique. Since the said stability and uniformity determination method and procedure is suitable in in terms of accuracy, the test setup performance is described by the uncertainty of measurements where the testing conditions are affected by the given parameters: (1) environmental conditions of ambient temperature ranging from 20 to 26°C and relative humidity range of 40 to 55% R.H. , (2) a stirred-liquid and overflowing calibration liquid bath functioning at 150°C to 250°C must be the enclosure used and silicone oil as the only liquid utilized during the test and (3) test thermometers' accuracy of at least + 0.05°C.